\documentclass[11pt,a4paper]{article}
\usepackage{graphicx}
\usepackage{epstopdf}
\usepackage{amstext,amsfonts,amsbsy,amssymb,bbm}
\usepackage{amsmath}
\usepackage{pdfsync}
\usepackage{a4wide}
\usepackage{times}

\newcommand{\Img}{\mathop{\mbox{Img}}}
\newcommand{\Ker}{\mathop{\mbox{Ker}}}

\def\Tr{\mathop{\rm Tr}\nolimits}

\newcommand {\be}[1]{\begin{eqnarray} \mbox{$\label{#1}$}  }
\newcommand{\ee}{\end{eqnarray}}

\newcommand{\wt}[1]{\widetilde{#1}}
\newcommand{\rmi}{\mbox{$\rm i$}}

\begin{document}

 \bibliographystyle{unsrt}




 \title{Extremal states of positive partial transpose in a system of
   three qubits}

\author{{\O}yvind Steensgaard Garberg, B{\o}rge Irgens, and Jan Myrheim,\\
Department of Physics,
Norwegian University of Science and Technology,\\
N--7491 Trondheim, Norway}



\maketitle

\begin{abstract}
  We have studied mixed states in the system of three qubits with the
  property that all their partial transposes are positive, these are
  called PPT states.  We classify a PPT state by the ranks of the
  state itself and its three single partial transposes.  In random
  numerical searches we find entangled PPT states with a large variety
  of rank combinations.  For ranks equal to five or higher we find
  both extremal and nonextremal PPT states of nearly every rank
  combination, with the restriction that the square sum of the four
  ranks of an extremal PPT state can be at most 193.  We have studied
  especially the rank four entangled PPT states, which are found to
  have rank four for every partial transpose.  These states are all
  extremal, because of the previously known result that every PPT
  state of rank three or less is separable.  We find two distinct
  classes of rank 4444 entangled PPT states, identified by a real
  valued quadratic expression invariant under local
  $\mbox{SL}(2,\mathbbm{C})$ transformations, mathematically
  equivalent to Lorentz transformations.  This quadratic Lorentz
  invariant is nonzero for one class of states (type~I in our
  terminology) and zero for the other class (type~II).  The previously
  known states based on unextendible product bases is a nongeneric
  subclass of the type~I states.  We present analytical constructions
  of states of both types, general enough to reproduce all the rank
  4444 PPT states we have found numerically.  We can not exclude the
  possibility that there exist nongeneric rank four PPT states that we
  do not find in our random numerical searches.
\end{abstract}

{\small PACS numbers: 03.67.Mn, 02.40.Ft, 03.65.Ud}


\section{Introduction}

As one of the most fascinating features of quantum mechanics
entanglement has been intensively studied in the last
decades~\cite{RPMKHorodecki}.  The Bell inequalities~\cite{Bell64}
turned the philosophical discussion between Einstein and Bohr into a
subject for experimental
investigations~\cite{Freedman72,Aspect81,Aspect82a,Aspect82b}.  The
Bell inequalities are deduced from the hypothesis of local realism,
and apply to statistical correlations in composite systems with two
subsystems.  One basic weakness of all such experiments is the
inherent statistical uncertainties of the observed correlations.

The striking new feature of the experiments proposed by Greenberger,
Horne, and Zeilinger~\cite{GHZ89} and Mermin~\cite{Mermin90}, in
composite systems with three or more subsystems, is that the
correlations to be tested are absolute and no longer statistical.  One
single observation is sufficient to demonstrate that quantum mechanics
violates local realism.  Experiments of this kind have also been
made~\cite{Bouwmeester99}.

A central problem in the study of entanglement is how to determine,
theoretically or experimentally, whether a state in a composite system
made up of two or more components is entangled or separable.  The
answer is simple for a pure state: it is entangled if it is not a
tensor product.  The separability problem for mixed states, on the
other hand, has proven to be highly nontrivial~\cite{Gurvits03} and
does not yet have a solution which is satisfactory for practical use.
The simplest and best known condition for the separability of a mixed
state is the Peres criterion \cite{Peres96}, which states that a
separable state remains positive semidefinite under partial
transpositions.  This necessary condition is in general not
sufficient, but applying it requires very little computational effort
and the separability problem is therefore in essence reduced to
determining whether mixed states with positive partial transposes (PPT
states) are entangled or separable.

A partial transposition is a mathematical operation that transposes
one or more factors in a tensor product, for example,
\begin{equation}
(\rho_1\otimes\rho_2\otimes\rho_3)^{T_2}
=\rho_1\otimes\rho_2^{\,T}\otimes\rho_3\;.
\end{equation}
The point is that it is a well defined operation even for a matrix
which is not a tensor product (see
Appendix~\ref{sec:partialtranspositions}).  A Hermitian matrix
representing a mixed state of an $n$-partite system may appear in
$2^n$ different versions related by partial transpositions, but these
can be separated into two sets where every matrix in one set is the
total transpose of a matrix from the other set.  Since the total
transpose $T$ preserves eigenvalues, half of the $2^n$ partial
transposes are superfluous when we evaluate whether or not a mixed
state $\rho$ is a PPT state.

In the present article we consider PPT states in a system of three
qubits.  In this system it is sufficient to watch the eigenvalues of
$\rho, \rho^{T_1}, \rho^{T_2},$ and $\rho^{T_3}$, where $T_i$ is the
partial transposition with respect to subsystem $i$.  It is necessary
to require that all these four are positive, since it is quite
possible for three of them to be positive, while the fourth one is
not.  The four remaining partial transposes are obtained from these by
a total transposition.  For example, $T_1T_2 = T_3T$, because
$T=T_1T_2T_3$, partial transpositions commute, and $T_i^{\,2}$ is the
identity operation.

We write $\mathcal{D}$ for the set of all unnormalized mixed states,
and $\mathcal{D}_1$ for the set of all mixed states normalized to unit
trace.  Similarly, we write $\mathcal{S}$ or $\mathcal{S}_1$ for the
set of all separable states, and $\mathcal{P}$ or $\mathcal{P}_1$ for
the set of all PPT states, unnormalized or normalized.  All of these
are convex sets.

An extremal point in a convex set is a point which is not a convex
combination of other points in the set.  The sets of normalized
states, $\mathcal{D}_1$, $\mathcal{S}_1$, and $\mathcal{P}_1$ are
compact and hence completely described by their extremal points, in
the sense that all nonextremal points may be written as convex
combinations of the extremal points.  The extremal points of
$\mathcal{D}_1$ are the pure states, and the extremal points of
$\mathcal{S}_1$ are the pure product states.  The inclusions
\begin{equation}
\mathcal{S}_1\subset\mathcal{P}_1\subset\mathcal{D}_1\;,
\end{equation}
together with the fact that the pure product states are extremal
points in both $\mathcal{S}_1$ and $\mathcal{D}_1$, imply that the
pure product states are also extremal points of $\mathcal{P}_1$.  It
is easy to prove that all pure nonproduct states are not in
$\mathcal{P}_1$ and therefore entangled.  The pure product states are
the only PPT states of rank one, since only pure states have rank one.
Because $\mathcal{S}_1$ and $\mathcal{P}_1$ are not identical in the
three qubit system, $\mathcal{P}_1$ must have extremal points of rank
higher than one giving rise to all the entangled PPT states.  These
extremal entangled PPT states are almost completely unknown, and that
situation motivates our study presented here.

Previous studies of multipartite entanglement have been mostly
concerned with pure states.  D{\"u}r, Vidal, and Cirac~\cite{Dur00}
classified pure states in the three qubit system into six equivalence
classes, based on the type of entanglement possessed by a state.  A
pure state is of the form $\psi\psi^{\dag}$ with
$\psi\in\mathbbm{C}^8$.  Two (unnormalized) vectors $\psi$ and $\phi$
are considered equivalent if
\begin{equation}
\phi=(V_1 \otimes V_2 \otimes  V_3)\,\psi
\end{equation}
with $V_i\in\mbox{SL}(2,\mathbbm{C})$.  One class contains the
separable (unentangled) pure states where $\psi$ is a $2\times 2\times
2$ dimensional product vector.  Three other classes contain the
biseparable pure states where $\psi$ is a product vector in one of the
three splittings into one system of one qubit and one system of two
qubits.  All states in these three classes have only bipartite
entanglement.  The last two classes contain states with two
inequivalent types of genuine tripartite entanglement.  There is the W
class, exemplified by the unnormalized state, in Dirac notation,
\begin{equation}
|\,\mbox{W}\,\rangle=|100\rangle+|010\rangle+|001\rangle\;.
\end{equation}
Finally there is the GHZ class, exemplified by the
Greenberger--Horne--Zeilinger state
\begin{equation}
|\,\mbox{GHZ}\,\rangle=|000\rangle+|111\rangle\;.
\end{equation}
In matrix notation, with
\begin{equation}
|0\rangle\;\leftrightarrow\;\begin{pmatrix}1\\0\end{pmatrix}\;,\qquad
|1\rangle\;\leftrightarrow\;\begin{pmatrix}0\\1\end{pmatrix}\;,
\end{equation}
the W and GHZ states are
\begin{equation}
\psi_{\mbox{\small W}}
=\begin{pmatrix}0\\1\\1\\0\\1\\0\\0\\0\end{pmatrix}\;,\qquad
\psi_{\mbox{\small GHZ}}
=\begin{pmatrix}1\\0\\0\\0\\0\\0\\0\\1\end{pmatrix}\;.
\end{equation}

The corresponding classification scheme for mixed states in the three
qubit system was introduced by Ac\'{\i}n et al.~\cite{Acin2001}.  The
scheme involves four convex sets with an onion structure where each
set is defined by including increasingly larger sets of pure states.
The innermost set is the set $\mathcal{S}$ of all separable states,
consisting of all states that are convex combinations of pure product
states.  The second set $\mathcal{B}$ includes all the biseparable
pure states.  The third set $\mathcal{W}$ includes also pure states
with W entanglement.  The fourth set includes the last remaining class
of pure states, the GHZ entangled states.  Because it includes all
pure states the fourth set is $\mathcal{D}$, the set of all mixed
states.  The authors conjecture that all entangled PPT states are
members of the third set $\mathcal{W}$.

Karnas and Lewenstein~\cite{Karnas2001} proved that all states of
ranks two and three in the three qubit system are separable.  They
applied the range criterion of entanglement for the three qubit system
as a special case of the $2\times 2\times N$ dimensional system.
Bennett et al.~\cite{Bennett99,DiVincenzo03} introduced the PPT states
based on unextendible product bases (UPBs) as examples of entangled
PPT states in the three qubit system.  These three qubit UPB states
have rank four.  They are entangled because there is no product vector
in the range of a UPB state, and they are extremal PPT states because
no entangled states of lower rank exist.


\subsubsection*{Sorting states into equivalence classes}

We want to classify the three qubit states into
$\mbox{SL}\otimes\mbox{SL}\otimes\mbox{SL}$ equivalence classes,
defining two unnormalized density matrices $\rho$ and $\sigma$ to be
equivalent if
\begin{equation}
\label{eq:VVVrhoVVV}
\sigma
=(V_1 \otimes V_2 \otimes  V_3)\,\rho\,
(V_1 \otimes V_2 \otimes V_3)^{\dag}\;,
\end{equation}
with $V_i\in\mbox{SL}(2,\mathbbm{C})$.  This definition is useful
because equivalent density matrices have the same entanglement
properties, although quantitative measures of entanglement will be
different.  Qualitative properties will be the same, such as tensor
product structure of pure states, decomposition of mixed states as
convex combinations of pure states, rank and positivity of states and
all their partial transposes, and so on.

The relation between the group SL(2,$\mathbbm{C}$) and the group of
continuous Lorentz transformations is well known, and is reviewed here
in Appendix~\ref{sec:SL2C}.  From a density matrix in the three qubit
system we define one quadratic and four quartic real Lorentz
invariants, so called because they are invariant under
$\mbox{SL}\otimes\mbox{SL}\otimes\mbox{SL}$ transformations as in
equation~(\ref{eq:VVVrhoVVV}).  They are also invariant under partial
transpositions, because a partial transposition may be interpreted as
a parity transformation, which is a discrete Lorentz transformation.

This means, for example, that the ratio between one quartic Lorentz
invariant and the square of the quadratic invariant will have the same
value for all the states in one equivalence class and all their
partial transposes.  Taking the ratio between Lorentz invariants is
necessary in order to cancel out any normalization factor in the
density matrix.  If two density matrices are not in the same
equivalence class, their inequivalence will most likely be revealed
when we calculate their invariants.

\subsubsection*{Outline of this article}

We have investigated both extremal and nonextremal PPT states in the
system of three qubits using both numerical and analytical methods.
We classify the states according to the ranks $m_0,m_1,m_2,m_3$ of
$\rho, \rho^{T_1}, \rho^{T_2},$ and $\rho^{T_3}$.

In Section~\ref{sec:Nummethods} we present the two main numerical
algorithms that we have used in random searches for PPT states.  We
have searched systematically for extremal states of unspecified ranks,
and also for states of specified ranks that are not necessarily
extremal.

In Section~\ref{sec:Results} we present results from the searches.  We
find PPT states with a wide variety of rank combinations.  The rank
1111 states are the pure states.  The states of ranks 2222 and 3333
are all separable, in agreement with previously known results.  An
interesting class of extremal PPT states are those of rank 4444, they
contain genuine tripartite entanglement, since they are separable in
any bipartite splitting of the three qubit system.  We have studied
these states in detail and shown how to construct them analytically,
as reported in Section~\ref{sec:rank4444}.

All the other states we find have all ranks equal to five or higher.
We find PPT states, extremal or nonextremal, of very nearly every rank
combination.  For an extremal PPT state the square sum of ranks can be
at most 193.  Up to this upper limit we also find extremal PPT states
of very nearly every rank combination.

In Section~\ref{sec:rank4444} we present our understanding of the
rank~4444 extremal PPT states.  It turns out that the key to
understanding them is the fact that they are biseparable in three
different ways~\cite{Kraus2000}.  There are two distinct classes of
such states, we call them simply type~I and type~II.  The most obvious
distinction is that the quadratic Lorentz invariant is nonzero for
states of type~I and zero for states of type~II.

When we sort these states further into
$\mbox{SL}\otimes\mbox{SL}\otimes\mbox{SL}$ equivalence classes, we
find that every equivalence class of type~I states contains a density
matrix which is real and symmetric under all partial transposistions.
We define this matrix to be a standard form for all the states in the
equivalence class.  The UPB states are a subclass of the type~I
states, but they are not sufficiently generic that we find any of them
in our random searches.

It is straightforward to construct the most general density matrix
which is biseparable, real and symmetric under all partial
transpositions.  The empirical result of our numerical searches is
that this construction reproduces all the rank 4444 PPT states of type~I.
The $\mbox{SL}\otimes\mbox{SL}\otimes\mbox{SL}$ equivalence classes
are parametrized by seven continuous real parameters.

A rank 4444 PPT state of type~II can not in general be transformed to
a real form.  We have studied our numerical examples of such states
and observed several special properties that they have.  Based on
these observations we present an analytical construction general
enough to reproduce all the type~II states found numerically.  The
$\mbox{SL}\otimes\mbox{SL}\otimes\mbox{SL}$ equivalence classes in
this case are parametrized by one continuous complex parameter.

Our work presented here extends previous studies of PPT states in
bipartite composite systems
\cite{LeinaasMyrheim06,LeinaasMyrheim07,Sollid10a,
  Sollid10b,SollidLeinaas10,Sollid11}.
There are obvious similarities and differences, especially between
the $3\times 3$ and the $2\times 2\times 2$ systems.


\section{Numerical methods}
\label{sec:Nummethods}

Most of the PPT states that we have studied numerically were found by
means of two algorithms that search for extremal PPT states or for PPT
states with a specified combination of ranks.  In this section we
present briefly these two main algorithms.

\subsection{Random search for extremal PPT states}

Extremal states in $\mathcal{P}_1$, the convex set of PPT states of
unit trace, are found by an iterative process where we start with any
PPT state, pick a random search direction restricted to the unique
face where this state is an interior point, and follow this direction
to the edge of the face, which is a face of lower dimension.  The next
search direction is restricted to this new face.  The procedure is
repeated until a face of dimension zero, which is an extremal point of
$\mathcal{P}_1$, is located.  An extremal point of $\mathcal{P}_1$
defines an extremal ray, a one dimensional face, of $\mathcal{P}$, the
cone of unnormalized PPT states.

A valid search direction from a given PPT state $\rho\in\mathcal{P}_1$
is a nonzero matrix $\sigma$ such that
$\rho+\epsilon\sigma\in\mathcal{P}_1$ for every $\epsilon$ in some
finite interval, $\epsilon_1\leq\epsilon\leq\epsilon_2$, with
$\epsilon_1<0<\epsilon_2$.  This means that $\sigma$ is a traceless
Hermitian matrix satisfying the four equations
\begin{equation}
P_i\sigma^{T_i} P_i=\sigma^{T_i}\;,\qquad i=0,1,2,3\;,
\label{eq:PQ}
\end{equation}
where $P_i$ is the orthogonal projection onto $\Img\rho^{T_i}$, and
where we write $\rho^{T_0}=\rho$.  The real vector space $H$ of
Hermitian $N\times N$ matrices has dimension $N^2$ and is a Hilbert
space with the scalar product
\begin{equation}
\langle A,B\rangle = \langle B,A\rangle = \Tr(AB)\;.
\end{equation}
The constraints in equation~\eqref{eq:PQ} can now be written as
eigenvalue equations $\mathbf{P}_i\sigma=\sigma$ for linear projection
operators $\mathbf{P}_i$ on $H$, symmetric with respect to this scalar
product, defined as follows,
\begin{equation}
\mathbf{P}_i\sigma=(P_i\sigma^{T_i} P_i)^{T_i}\;,
\qquad i=0,1,2,3\;.
\label{eq:PQ2}
\end{equation}
The four eigenvalue equations may be combined into one single
eigenvalue equation
\begin{equation}
\displaystyle\sum\limits_{i=0}^3 \mathbf{P}_i\,\sigma=4\sigma\;.
\label{eq:PPP}
\end{equation}
This equation always has the trivial solution $\sigma=\rho$, but
$\rho$ is not traceless.  If $\rho$ is the only solution it is an
extremal point of $\mathcal{P}_1$.  The number of linearly independent
traceless solutions to this eigenvalue equation is the dimension of
the restricting face of $\mathcal{P}_1$.  We find a random search
direction by finding a complete set of eigenvectors solving
equation~\eqref{eq:PPP}, then choosing $\sigma_0$ as a random linear
combination of these, and finally obtaining a traceless solution as
\begin{equation}
\sigma = \sigma_0 - (\Tr\sigma_0)\,\rho\;.
\end{equation}
The bipartite version of this algorithm is described in more detail
in~\cite{LeinaasMyrheim07}.

\subsubsection{Separability test for low rank states}
\label{sec:separabilitytest}

A modified version of this algorithm can be used as a separability
test for low rank states.  If a PPT state is not extremal it can be
written as a convex combination of extremal PPT states of lower rank.
A set of such states can be found by searching in both directions
$\sigma$ and $-\sigma$ every time a search direction $\sigma$ is
chosen.  The original state is separable if all the extremal states
found in this process are pure.  In general it may be possible to
write a separable state as a convex combination involving some mixed
extremal states.  However, for low rank states few mixed extremal
states of lower rank are available and these ``false negatives'' are
therfore less likely to occur.

\subsubsection{Upper limit on the ranks}

We derive an upper limit on the ranks of extremal PPT states by
counting the number of constraints on $\sigma$. Since each operator
$\mathbf{P}_i$ is a projection with eigenvalues 0 and 1, the
constraint equation $\mathbf{P}_i\sigma=\sigma$ places a number of
constraints on $\sigma$ equal to the dimension of the kernel of
$\mathbf{P}_i$.  If $\rho^{T_i}$ has rank $m_i$, then the rank of
$\mathbf{P}_i$ is $m_i^2$, and the dimension of its kernel is
$N^2-m_i^2$.  Thus, for a given $\rho$ the total number of constraints
on $\sigma$, apart from the zero trace condition, is
\begin{equation}
4N^2 -\displaystyle\sum\limits_{i} m_i^2\;.
\end{equation}
These constraints need not be linearly independent, hence this is an
upper limit on the number of linearly independent constraints.  Since
$\rho$ is an extremal PPT state if and only if it is the only
Hermitian matrix satisfying the constraints, the number of linearly
independent constraints when $\rho$ is extremal must be $N^2-1$.  The
total number of constraints must be at least as large.  This implies
the following limit on the ranks of extremal PPT states and their
partial transposes in a three qubit system,
\begin{equation}
\displaystyle\sum\limits_{i} m_i^2 \leq 3N^2+1\;.
\label{eq:limit}
\end{equation}

The derivation of the formula involves a sum over the independent
partial transposes, the number of which is $2^{n-1}$ for an
$n$-partite system.  The formula can therefore be generalized to
\begin{equation}
\displaystyle\sum\limits_{i} m_i^2 \leq (2^{n-1}-1)N^2+1
\label{eq:limit2}
\end{equation}
for an $n$-partite system of total dimension $N$.  For large $n$ this
inequality is a very mild restriction on the ranks.

\subsection{Search for PPT states of specified low ranks}

The second algorithm we describe take as input the desired ranks
$m_0,m_1,m_2,m_3$ of $\rho$ and its partial transposes.  After we find
a $\rho$ with the specified ranks we may check, by the algorithm just
described, whether or not it is extremal.

The equations defining a Hermitian matrix $\rho$ as a PPT state of the
given ranks are of the form $\mu=0$, where $\mu$ is a list of all the
lowest eigenvalues of $\rho$ and its partial transposes.  We include
the $N-m_i$ lowest eigenvalues of $\rho^{T_i}$ (recall our definition
$\rho^{T_0}=\rho$).
When we write $\rho$ as a linear combination
\begin{equation}
\rho=\displaystyle\sum\limits_{j}x_j M_j
\end{equation}
of matrices $M_j$ forming a basis for the real vector space of
Hermitian matrices, the equations are of the form $\mu_i(x)=0$.  There
are $N^2$ real variables $x_j$, and the number of equations is
\begin{equation}
N_e=4N-\displaystyle\sum\limits_{i}m_i\;.
\end{equation}
%

\subsubsection{A minimization problem}

One way to solve the equations $\mu_i(x)=0$ is to minimize the
function
\begin{equation}
  f(x)=\displaystyle\sum\limits_{i}(\mu_i(x))^2
\end{equation}
and obtain a minimum value of zero.  We may use a random search
algorithm for finding the minimum.  This approach may be good enough
for many purposes, although not very efficient, and it has the
advantage of being easy to program.

Note that a minimum point of a function is in general not very
precisely determined numerically, because the function varies
quadratically around its minimum.  This problem does not arise here
because the desired minimal value is zero.

\subsubsection{The conjugate gradient method}

A more refined and efficient iterative approach is to linearize the
equations and solve in each iteration the approximate, linearized
equations using the conjugate gradient method.  A similar algorithm
for the bipartite case is described in~\cite{Sollid10a}.

Given an approximate solution $x$ we try to find a better solution
$x+\Delta x$ using the linearized equation
\begin{equation}
\mu(x+\Delta x)=\mu(x)+B(x)\,\Delta x=0\;,
\end{equation}
where the matrix elements of $B(x)$ are
$B_{ij}(x)=\partial \mu_i(x)/\partial x_j$.  We multiply this equation
by $B^T$, and get another equation
\begin{equation}
A\,\Delta x=b\;,
\end{equation}
where $A=B^TB$ and $b=-B^T\mu$.  The matrix $A$ is real and symmetric
as well as positive semidefinite.  It is likely to be singular, but
the last equation may anyway be solved by the conjugate gradient
method.

We calculate the components of $B$ using first order perturbation
theory.  If for example $\mu_i$ is the eigenvalue $\lambda_k$ of
$\rho$ then we use the formula
\begin{equation}
\frac{\partial \mu_i}{\partial x_j}
=\frac{\partial \lambda_k}{\partial x_j}
=\psi_k^\dagger\,\frac{\partial \rho}{\partial x_j}\,\psi_k
=\psi_k^\dagger M_j \psi_k\;.
\end{equation}
Here $\psi_k$ is the eigenvector corresponding to the eigenvalue
$\lambda_k$ of $\rho$.  This formula is based on nondegenerate
perturbation theory, which is strictly speaking not valid in the
present case where we try to make many eigenvalues simultaneously
equal to zero.  The formula nevertheless works well in practice.  We
use similar formulas for the eigenvalues of the partial transposes of
$\rho$.

\section{Results of numerical searches}
\label{sec:Results}

\subsection{Ranks of PPT states found numerically}

Numerical searches for PPT states of specified ranks were successful
for most of the rank combinations.  Note however that our search
algorithm might occasionally produce PPT states of lower ranks than
the ones specified.  Note also that for the purpose of classification
of states there is full symmetry between $\rho$ and all its seven
partial transposes (including the double partial transposes and the
total transpose).  Furthermore, with three identical subsystems there
is full symmetry between states related by an interchange of the
subsystems.  Therefore when we list the ranks $m_0,m_1,m_2,m_3$ we use
the convention that $m_0\leq m_1\leq m_2\leq m_3$.

The PPT states found with specified ranks might be separable or
entangled, extremal or nonextremal.  We used them as starting points
in searches for extremal PPT states.  This would either show them to
be extremal, or return extremal PPT states of lower ranks.

A list of all confirmed rank combinations and the corresponding square
sums of ranks is given in Table~\ref{tab:results}.  By
equation~\eqref{eq:limit}, for an extremal PPT state the square sum of
ranks can not be larger than 193.  We found no PPT states at all with
the very asymmetric rank combinations 5568, 5588, or 5888.  Otherwise,
the rank combinations $2222$, $3333$, and $5688$ are the only cases
where the inequality allows extremal PPT states to exist, but we did
not find any.

\begin{table}[h]
\centering
\begin{tabular}{|lr|lr|lr|}
\hline
1111 &  4&5666 &133&6677&170\\
2222*& 16&5667 &146&6678&185\\
3333*& 36&5668 &161&6688&200\\
4444 & 64&5677 &159&6777&183\\
5555 &100&5678 &174&6778&198\\
5556 &111&5688*&189&6788&213\\
5557 &124&5777 &172&6888&228\\
5558 &139&5778 &187&7777&196\\
5566 &122&5788 &202&7778&211\\
5567 &135&6666 &144&7788&226\\
5577 &148&6667 &157&7888&241\\
5578 &163&6668 &172&8888&256\\
\hline
\end{tabular}
\caption{A list of all confirmed rank combinations and
the corresponding sums of squared ranks.
The upper limit on the sum of squared ranks
of extremal PPT states, as given by
equation~\eqref{eq:limit}, is 193.
There are three cases, each marked by an asterisk,
where this limit allows extremal states to exist
but only nonextremal states were found.}
\label{tab:results}
\end{table}

A similar study on the ranks of extremal states in low dimensional
bipartite systems is presented in \cite{Sollid10a}. The structure of
the rank combinations of our three qubit states closely resembles the
structure in the 3$\times$3, 3$\times$4, 3$\times$5, and 4$\times$4
systems. For the lowest rank combinations only states where all
partial transposes have the same rank are found, and below some
threshold rank, four in our case, all these states are
separable. Above the threshold extremal states are found for
essentially all rank combinations where PPT states exist and where the
upper limit given in equation~\eqref{eq:limit} holds. The sole
exception is the rank combination 5688 where extremal states are
allowed by the limit, but we were unable to confirm their existence.
We consider it likely that extremal states of this rank combination do
exist, but are difficult to find numerically by our methods.

\subsection{Rank 4444 PPT states}

The rank four extremal PPT states are of special interest because they
are the entangled PPT states of lowest rank.  Thus, any rank four PPT
state is either extremal or separable.  Another special property of
the rank four PPT states is that all their partial transposes have the
same rank, we find no PPT states of rank 4445, for example.


Nine out of 196 extremal rank four states found in a random numerical
search stand out because their quadratic Lorentz invariant is zero,
which means in practice of order $10^{-14}$ to $10^{-16}$.
We find a qualitative difference between the states with nonzero
quadratic invariant and those with the invariant equal to zero.  In
particular, in our numerical sample we see no continuous transition
from invariants of order one to those of order $10^{-14}$.
 
We will discuss the rank 4444 entangled PPT states in much more detail
in Section~\ref{sec:rank4444}.

\subsection{PPT states with special symmetries}

An obvious way to construct a PPT state is to construct a positive
Hermitian matrix which is symmetric under all three partial
transpositions $T_1,T_2,T_3$.  These symmetries together imply
symmetry under the total transposition $T=T_1T_2T_3$, which makes the
matrix real.  The most general Hermitian matrix symmetric under all
three partial transpositions contains 27 real parameters.  See
Appendix~\ref{sec:partialtranspositions}.

Completely symmetric PPT states of rank eight are easily generated as
follows.  Generate a random complex matrix, and multiply it with its
Hermitian conjugate to obtain a Hermitian and positive matrix.  Add to
this matrix all its seven partial transposes.  The resulting matrix is
completely symmetric and has a high probability of being positive.

Another way to generate a completely symmetric PPT state of any rank
$m$ is the following.  Start with any matrix in the 27 dimensional
space of completely symmetric matrices.  Then take the square sum of
its $8-m$ lowest eigenvalues, and minimize this by varying the matrix
so as to obtain the minimum value of zero.  We used this method to
create states of rank four, five, six, and seven. The rank four states
were all extremal, while all the higher rank states were nonextremal.

It is also possible to relax the symmetry requirements and only
require symmetry under two partial transpositions, say $T_1$ and
$T_2$.  This raises the number of free parameters in a Hermitian
matrix from 27 to 36 and allows it to be complex.  A state $\rho$ with
the symmetries $\rho=\rho^{T_1}=\rho^{T_2}$ is still guaranteed to be
a PPT state, because $\rho^{T_3}=\rho^T$ has the same eigenvalues as
$\rho$.  Only about 75\% of the rank four states generated with these
symmetries were extremal.


\subsection{Decomposition}

The separability test described in
Subsection~\ref{sec:separabilitytest} is of particular interest for
the nonextremal rank 5555 states, which are the states of lowest rank
where the test is nontrivial.  Among the states of this kind found in
random searches for states of specified ranks we found both separable
and entangled states.  Every entangled state was found to be an
interior point of a one dimensional face of $\mathcal{P}_1$, this face
is then a line segment with a pure state at one end and a rank 4444
extremal PPT state at the other end.  Every separable state was found
to be an interior point of a four dimensional face of $\mathcal{P}_1$,
as expected when it is a convex combination of five pure states.

We also tested the rank five states that were constructed so as to be
symmetric under all partial transpositions.  The result is that every
such state is an interior point of a face of dimension five, and
decomposes always into two rank four extremal states having the same
symmetry.

We found the last result surprising, and therefore applied the same
test to symmetric states with higher ranks.  The rank six states
behave in a very similar manner, decomposing into four symmetric rank
four states that are always extremal.  The rank seven states, on the
other hand, decompose mainly into rank 6677 states which obviously do
not possess the same symmetry.  Every rank eight state decomposes into
32 rank 6777 states that also do not possess any symmetries.

\section{Rank four entangled PPT states as biseparable states}
\label{sec:rank4444}

We find empirically that the entangled and extremal rank 4444 PPT
states in dimension $2\times 2\times 2$ fall into two qualitatively
different classes, which we choose to call types I and II.  By
definition, a state $\rho$ of type~I has a nonzero (always positive)
value of the quadratic Lorentz invariant
\begin{equation}
\rho^{\lambda\mu\nu}\rho_{\lambda\mu\nu}
=-\frac{1}{8}\Tr(\rho^TE\rho E)\;,
\end{equation}
whereas this invariant vanishes for a state $\rho$ of type~II.  See
Appendix~\ref{sec:SL2C}.  As we shall see, another characteristic
difference is that a state of type~I can be transformed by a product
transformation to a standard form where it is real and symmetric under
all partial transpositions, whereas a state of type~II can not in
general be transformed to a real form.  Type~I includes as a special
case the so called UPB states constructed from unextendible product
bases (UPBs), introduced in~\cite{Bennett99,DiVincenzo03}, which we will also
describe here.

The basic fact leading to an understanding of these states is that if
$\rho$ is a density matrix of rank four in dimension $2\times 2\times
2$, and if $\rho^{T_1}\geq 0$, then $\rho$ is separable in dimension
$2\times 4$.  Similar results hold of course for the cases
$\rho^{T_2}\geq 0$ and $\rho^{T_3}\geq 0$.  This has been proved
analytically~\cite{Kraus2000} and also verified
numerically~\cite{Sollid10a}.  We will describe in this section how to
use the threefold biseparability in order to understand our rank 4444
PPT states.

We first discuss some common properties of the two types of states,
and a new method for constructing them numerically, before we turn to
the more detailed mathematical understanding of the two types
separately, including the UPB states.

\subsection{General considerations}
\label{subsec:generalconsiderations}

Consider a generic four dimensional subspace
$\mathcal{U}\subset\mathbbm{C}^8$.  It contains exactly four product
vectors 
\begin{equation}
e_i=x_i\otimes u_i\;,\qquad i=1,2,3,4\;,
\end{equation}
of dimension $2\times 4$, which form a basis for $\mathcal{U}$, and
which may be calculated by the method described in
Appendix~\ref{sec:prodvecinsubspace}.  It contains also exactly four
product vectors
\begin{equation}
f_j=y_j\otimes_s v_j\;,\qquad j=1,2,3,4\;,
\end{equation}
where $y_j\in\mathbbm{C}^2$, $v_j\in\mathbbm{C}^4$, and where we use
the split tensor product defined in
Appendix~\ref{sec:splittensorproduct}.  These product vectors form
another basis for $\mathcal{U}$.  And it contains exactly four product
vectors
\begin{equation}
g_k=w_k\otimes z_k\;,\qquad k=1,2,3,4\;,
\end{equation}
of dimension $4\times 2$, forming a third basis for $\mathcal{U}$.

A density matrix $\rho$ of rank four with $\Img\rho=\mathcal{U}$ and
$\rho^{T_1}\geq 0$ must be biseparable in dimension $2\times 4$, and
hence of the form
\begin{equation}
\label{eq:bisep1}
\rho=\sum_{i=1}^4\lambda_i\,e_ie_i^{\dag}
\quad\mbox{with all}\quad\lambda_i>0\;.
\end{equation}
Note that although $\rho$ is separable as a bipartite state, it is
necessarily entangled as a tripartite state, except in the special
case when the four vectors $e_i$ are $2\times 2\times 2$ product
vectors, which means that they are identical to the vectors $f_j$, and
also identical to the vectors $g_k$.

If $\rho^{T_2}\geq 0$, then $\rho$ must be biseparable by the split
tensor product, and we must have
\begin{equation}
\label{eq:bisep2}
\rho=\sum_{j=1}^4\mu_j\,f_jf_j^{\dag}
\quad\mbox{with all}\quad\mu_j>0\;.
\end{equation}
If $\rho^{T_3}\geq 0$, then $\rho$ must be biseparable in dimension
$4\times 2$, and
\begin{equation}
\label{eq:bisep3}
\rho=\sum_{k=1}^4\nu_k\,g_kg_k^{\dag}
\quad\mbox{with all}\quad\nu_k>0\;.
\end{equation}

\subsubsection{Constructing PPT states numerically as biseparable states}
\label{subsubsec:PPTstatesbisep}

We find that with a generic subspace $\mathcal{U}$ any two of the
three equations~(\ref{eq:bisep1}), (\ref{eq:bisep2}),
and~(\ref{eq:bisep3}) are incompatible, they have no common nonzero
solution for $\rho$.  If we require that two of the three equations,
for example~(\ref{eq:bisep1}) and~(\ref{eq:bisep2}), should be
compatible, then this restricts the subspace $\mathcal{U}$.  It is
easy to find subspaces where two of the three equations are
compatible, but then the third equation will in general be
incompatible with the two.  This means that we may have, for example,
$\rho\geq 0$, $\rho^{T_1}\geq 0$, and $\rho^{T_2}\geq 0$, but
$\rho^{T_3}\not\geq 0$.

When we want a PPT state with
$\rho\geq 0$ and $\rho^{T_i}\geq 0$ for $i=1,2,3$, we need to find a
subspace $\mathcal{U}$ where the equations~(\ref{eq:bisep1})
and~(\ref{eq:bisep2}) are compatible, and at the same time the
equations~(\ref{eq:bisep1}) and~(\ref{eq:bisep3}).  We will describe
next how to solve these compatibility problems numerically.

The equations~(\ref{eq:bisep1}) and~(\ref{eq:bisep2}), apart from the
positivity conditions on the coefficients $\lambda_i$ and $\mu_j$, are
compatible when there is at least one linear dependence between the
eight Hermitian $8\times 8$ matrices $e_ie_i^{\dag}$ and
$f_jf_j^{\dag}$.  We do the calculation numerically by converting the
eight matrices to eight vectors in $\mathbbm{R}^{64}$ and making a
singular value decomposition of the $64\times 8$ matrix.  The eight
singular values are non-negative, by definition, and we want the
smallest one to be zero.  This is a minimization problem where we vary
the subspace $\mathcal{U}$ in order to minimize the smallest singular
value.  We solve this in practice by a random search method.

An output of the singular value decomposition, after minimization, is
the two sets of coefficients $\lambda_i$ and $\mu_j$ corresponding to
the singular value zero.  However, the singular value decomposition
puts no restriction on the signs of the coefficients $\lambda_i$ and
$\mu_j$.  If we get a solution with all $\lambda_i<0$, then we simply
switch all four signs in order to get all $\lambda_i>0$.  If, on the
other hand, we get four coefficients $\lambda_i$ with both signs, then
we have a matrix $\rho$ which is biseparable in two ways, but we can
not make it positive semidefinite.  This is then a subspace
$\mathcal{U}$ which we can not use.

The procedure for making the equations~(\ref{eq:bisep1})
and~(\ref{eq:bisep3}) compatible is the same, we formulate it as a
second minimization problem for a smallest singular value.  In order
to obtain a subspace $\mathcal{U}$ solving both problems
simultaneously we simply minimize the sum of the smallest singular
value from the first problem and the smallest singular value from the
second problem.

The numerical procedure described here for constructing PPT states of
rank four works well, although the random search for a minimum may
take some time.  In practice, we always get one singular value zero
with a corresponding unique solution for the coefficients $\lambda_i$,
$\mu_j$, and $\nu_k$.  However, it seems that each coefficient
$\lambda_i$ for $i=1,2,3,4$ comes with an essentially random sign.  We
need the same sign for all four coefficients in order to satisfy the
positivity condition $\rho\geq 0$, and on the average we have to try
about eight times before we succeed once.

The majority of the rank four PPT states found numerically by this
method are of type~I, having a nonzero value of the quadratic
invariant.  We find however also a significant fraction of states of
type~II, with vanishing quadratic invariant.  As already remarked,
these are two distinct classes of rank four PPT states.  We will
proceed next to discuss them separately, after discussing the so
called UPB states.

\subsubsection{Standard form}

In the generic case there always exists a product transformation
\begin{equation}
V=V_1\otimes V_2\otimes V_3\;,
\end{equation}
uniquely defined up to normalization, transforming the two dimensional
vectors $x,y,z$ into $V_1x$, $V_2y$, $V_3z$ such that the transformed
vectors after suitable normalizations have the standard form defined
in Appendix~\ref{sec:standardforms}, equation~(\ref{eq:standform1}),
with complex parameters $t_1,t_2,t_3$,
\begin{equation}
\label{eq:standformxyz}
x=\begin{pmatrix}1&0&1&t_1\\ 0&1&-1&1\end{pmatrix},\qquad
y=\begin{pmatrix}1&0&1&t_2\\ 0&1&-1&1\end{pmatrix},\qquad
z=\begin{pmatrix}1&0&1&t_3\\ 0&1&-1&1\end{pmatrix}.
\end{equation}
If $\rho$ is a density matrix of rank four with
$\Img\rho=\mathcal{U}$, then the product transformation $V$ transforms
$\rho$ into $V\rho V^{\dag}$, which in our terminology is
$\mbox{SL}\otimes\mbox{SL}\otimes\mbox{SL}$ equivalent to $\rho$.
This transformed matrix will then be a standard form for $\rho$.

Note that the standard form is not completely unique, because it
depends on the ordering within each of the three sets of product
vectors $e_i$, $f_j$, and $g_k$.  Thus there is a discrete ambiguity.

\subsection{States constructed from unextendible product bases}

The construction of rank four entangled PPT states in dimension
$2\times 2\times 2$ from unextendible product bases (UPBs) is due to
Bennett et al.~\cite{Bennett99,DiVincenzo03}.  It has been shown, both
numerically~\cite{Sollid10b,Sollid11} and
analytically~\cite{Skowronek11,Chen11}, that the UPB construction in
dimension $3\times 3$ produces all the rank four entangled PPT states
there.  In contrast, we find that the UPB states in dimension
$2\times 2\times 2$ are only a subclass of the rank four PPT states of
type~I, with nonvanishing quadratic invariant.

An unextendible product basis (a UPB) is a maximal set of orthogonal
product vectors which is not a complete basis for the vector space.
In other words, it is an orthogonal basis of product vectors for a
subspace, with the property that the orthogonal subspace contains no
product vectors.  As shown in Appendix~\ref{sec:UPBs} a UPB in
dimension $2\times 2\times 2$ consists of four product vectors
\begin{equation}
\psi_i=x_i\otimes y_i\otimes z_i\;,
\qquad
i=1,2,3,4\;.
\end{equation}
From these vectors, normalized to unit length, we construct a PPT
state $\rho$ as a normalized projection operator onto the orthogonal
subspace,
\begin{equation}
\label{eq:rhoUPB}
\rho = \frac{1}{4}\,
\bigg(\mathbbm{1}
 - \displaystyle\sum\limits_{i=1}^{4} \psi_i \psi_i^\dagger \bigg),
\end{equation}
where $\mathbbm{1}$ is the $8\times 8$ unit matrix.  This is obviously
a PPT state, since $\rho^{T_1}$, for example, is again an orthogonal
projection,
\begin{equation}
\rho^{T_1} = \frac{1}{4}\,
\bigg(\mathbbm{1}
 - \displaystyle\sum\limits_{i=1}^{4}
\widetilde{\psi}_i \widetilde{\psi}_i^\dagger \bigg)
\end{equation}
with
\begin{equation}
\widetilde{\psi}_i=x_i^{\ast}\otimes y_i\otimes z_i\;.
\end{equation}

The state $\rho$ constructed in this way is obviously entangled, since
$\Img\rho$ contains no product vector, and a characteristic property
of a separable state $\sigma$ is that $\Img\sigma$ is spanned by
product vectors.  From $\rho$ we get other rank four entangled PPT
states of the form $\widetilde{\rho}=aV\rho V^{\dag}$ where $V$ is a
product matrix, $V=V_1\otimes V_2\otimes V_3$, and $a$ is a
normalization constant.  We have then that
\begin{equation}
\Img\widetilde{\rho}=V\Img\rho\;,\qquad
\Ker\widetilde{\rho}=(V^{\dag})^{-1}\Ker\rho\;.
\end{equation}
The kernel of $\rho$ is spanned by the four orthogonal product vectors
$\psi_i$.  The kernel of $\widetilde{\rho}$ is spanned by the product
vectors
\begin{equation}
(V^{\dag})^{-1}\psi_i
=((V_1^{\dag})^{-1}x_i)\otimes
 ((V_2^{\dag})^{-1}y_i)\otimes
 ((V_3^{\dag})^{-1}z_i)\;,
\end{equation}
which are orthogonal if and only if $V$ is unitary.

The existence of a basis of product vectors in the kernel, orthogonal
or not, is the characteristic property of the rank four PPT states of
the UPB type and their $\mbox{SL}\otimes\mbox{SL}\otimes\mbox{SL}$
transforms.  These states are not generic, because a generic subspace
of dimension four contains no product vector.  In fact, a generic and
suitably normalized product vector
\begin{equation}
\psi=
\begin{pmatrix} 1\\ a\end{pmatrix}\otimes
\begin{pmatrix} 1\\ b\end{pmatrix}\otimes
\begin{pmatrix} 1\\ c\end{pmatrix}
\end{equation}
contains three complex parameters $a,b,c$.  In order to lie in a given
four dimensional subspace it has to satisfy $8-4=4$ linear
constraints, which it can not do in general with only three
parameters.

Using numerical methods, we have searched for product vectors in the
kernels of our extremal rank four PPT states, but found none.  We
conclude that there are no states equivalent to UPB states in our
numerical sample.  The UPB states exist, but are not sufficiently
generic to be easily found in random numerical searches.

The UPB has a standard form where all the vectors $x,y,z$ are real and
given by three continuous real parameters (three angles), as in
equation~(\ref{eq:standform4}).  Therefore $\rho$ has a standard form
in which it is symmetric under all partial transpositions,
$\rho=\rho^{T_1}=\rho^{T_2}=\rho^{T_3}$.

\subsection{Rank four PPT states with nonzero quadratic invariant}

We turn now from the UPB states to the more general class of rank 4444
extremal PPT states with nonvanishing quadratic invariant.  The key to
understanding these states comes from a study of our numerical
examples.  It turns out that all the states we have found numerically
may be transformed to a standard form where they are symmetric under
all partial transformations, and hence real.

Let $\rho$ be a state with nonvanishing quadratic invariant.  We
compute the three sets of four product vectors $e_i=x_i\otimes u_i$,
$f_j=y_j{\otimes}_sv_j$, and $g_k=w_k\otimes z_k$ in $\Img\rho$, with
$x_i,y_j,z_k\in\mathbbm{C}^2$ and $u_i,v_j,w_k\in\mathbbm{C}^4$, as
already defined in Subsection~\ref{subsec:generalconsiderations}.
Then we use a product transformation $V=V_1\otimes V_1\otimes V_3$ to
transform the two dimensional vectors $x,y,z$ to the standard form
given in equation~(\ref{eq:standformxyz}).

We find empirically that we always get real values for the parameters
$t_1,t_2,t_3$, and that the same transformation $V$ always makes $u,v,w$
real.  Moreover, the transformed matrix $V\rho V^{\dag}$ always
becomes symmetric under all partial transpositions.  These are the key
observations.  We have no mathematical proof that $\rho$ must have
these properties if its quadratic invariant is nonzero, but we take it
as a working hypothesis.  As such it is very powerful, and we will see
next how to construct the most general PPT state with these
properties.

\subsubsection{Explicit construction}

In equation~(\ref{eq:bisep1}) we may absorb a factor
$\sqrt{\lambda_i}$ into the vector $e_i=x_i\otimes u_i$ and write
\begin{equation}
\label{eq:bisep4}
\rho=\sum_{i=1}^4e_ie_i^{\dag}\;.
\end{equation}
The problem we face is to construct a density matrix $\rho$ which has
this form with $x$ and $u$ real, and is symmetric under all partial
transpositions, $\rho=\rho^{T_1}=\rho^{T_2}=\rho^{T_3}$.  We also want
$x$ to have the standard form given in
equation~(\ref{eq:standformxyz}), so that
\begin{equation}
e=\begin{pmatrix}u_1&0&u_3&t_1u_4\\ 0&u_2&-u_3&u_4\end{pmatrix}.
\end{equation}
Then $\rho$ has the following form,
\begin{equation}
\label{eq:standformrhoI}
\rho=\begin{pmatrix}A&B\\ B&C\end{pmatrix}
\end{equation}
with
\begin{eqnarray}
\label{eq:standformrhoII}
A&\!\!\!=&\!\!\!u_1u_1^T+u_3u_3^T+t_1^{\,2}\,u_4u_4^T\;,
\nonumber\\
B&\!\!\!=&\!\!\!-u_3u_3^T+t_1\,u_4u_4^T\;,
\\
C&\!\!\!=&\!\!\!u_2u_2^T+u_3u_3^T+u_4u_4^T\;.
\nonumber
\end{eqnarray}

Let us see what happens if we simply take $u$ to be a random
$4\times 4$ real matrix.  As shown in
Appendix~\ref{sec:partialtranspositions}, $\rho$ as given in
equation~(\ref{eq:standformrhoI}) is symmetric under all partial
transpositions if and only if the $4\times 4$ matrix $A$ is real and
has the following general form,
\begin{equation}
A=\begin{pmatrix}
a_1&a_5&a_6&a_7\\
a_5&a_2&a_7&a_8\\
a_6&a_7&a_3&a_9\\
a_7&a_8&a_9&a_4
\end{pmatrix},
\end{equation}
and $B$ and $C$ are also real and have similar forms.  In the case of
$A$ the only condition which is not automatically satisfied as a
consequence of equation~(\ref{eq:standformrhoII}) is that
$A_{41}=A_{32}$.  For $B$ and $C$ there are similar conditions.  In
the case of $B$ the condition takes the form
\begin{equation}
\label{eq:condB}
-u_{43}u_{13}+t_1u_{44}u_{14}=-u_{33}u_{23}+t_1u_{34}u_{24}\;.
\end{equation}
Since
\begin{equation}
u_i^T(\epsilon\otimes\epsilon)\,u_j
=u_{4i}u_{1j}-u_{3i}u_{2j}
-u_{2i}u_{3j}+u_{1i}u_{4j}\;,
\end{equation}
the condition that ensures the right form for $B$,
equation~(\ref{eq:condB}), may be rewritten as
\begin{equation}
-u_3^T(\epsilon\otimes\epsilon)\,u_3
+t_1\,u_4^T(\epsilon\otimes\epsilon)\,u_4
=0\;.
\end{equation}
We solve it simply by choosing
\begin{equation}
t_1=\frac{u_3^T(\epsilon\otimes\epsilon)\,u_3}
{u_4^T(\epsilon\otimes\epsilon)\,u_4}\;.
\end{equation}
Next, the condition that ensures the right form for $A$ is the
following,
\begin{equation}
\label{eq:condA}
 u_1^T(\epsilon\otimes\epsilon)\,u_1
+u_3^T(\epsilon\otimes\epsilon)\,u_3
+t_1^{\;2}\,u_4^T(\epsilon\otimes\epsilon)\,u_4
=0\;.
\end{equation}
Define now
\begin{equation}
\alpha^2=-
\frac{u_3^T(\epsilon\otimes\epsilon)\,u_3
+t_1^{\;2}\,u_4^T(\epsilon\otimes\epsilon)\,u_4}
{u_1^T(\epsilon\otimes\epsilon)\,u_1}\;.
\end{equation}
If $\alpha^2>0$, then we replace $u_1$ by $\alpha u_1$ and have a
solution of equation~(\ref{eq:condA}).  If instead $\alpha^2<0$, then
we change the sign of $u_1^T(\epsilon\otimes\epsilon)\,u_1$, for
example by changing the signs of the first two components of $u_1$.
With this new vector $u_1$ we have $\alpha^2>0$, and again we solve
equation~(\ref{eq:condA}) by substituting $\alpha u_1$ for $u_1$.

The condition that ensures the right form for $C$ is the
following,
\begin{equation}
\label{eq:condC}
 u_2^T(\epsilon\otimes\epsilon)\,u_2
+u_3^T(\epsilon\otimes\epsilon)\,u_3
+u_4^T(\epsilon\otimes\epsilon)\,u_4
=0\;.
\end{equation}
We solve it for $u_2$ in the same way as we solved
equation~(\ref{eq:condA}) for $u_1$.  This completes the construction
of $\rho$, except that we should normalize $\rho$ to have unit trace.

\subsubsection{Parameter counting}

A random $4\times 4$ real matrix $u$ contains 16 freely variable real
parameters.  In the process of constructing $\rho$ to be symmetric
under partial transformations we had to redefine the normalizations of
$u_1$ and $u_2$, this reduces the number of free parameters to 14.
The final normalization of $\rho$ reduces the number of parameters to
13.

Note that when we take $x$ to have its standard form and choose $u$
randomly, the resulting vectors $y$ and $z$ will most likely not have
their standard forms.  In order to transform them to standard forms we
will need two $\mbox{SL}(2,\mathbbm{R})$ transformations, containing
altogether six real parameters.  After we fix the standard forms of
$x,y,z$ there are no continuous degrees of freedom left in the product
transformation $V$.  There is however some discrete freedom left,
because we may permute the product vectors $e_i$, as well as the
vectors $f_j$ and the vectors $g_k$.

We conclude that there is a family of
$\mbox{SL}\otimes\mbox{SL}\otimes\mbox{SL}$ equivalence classes of
rank 4444 extremal PPT states with nonzero quadratic invariant
described by $13-6=7$ continuous real parameters.

\subsection{States with vanishing quadratic invariant}

In our numerical random searches for PPT states of rank four we found
a special class of rank 4444 PPT states with the property that the
quadratic Lorentz invariant of each state $\rho$ vanishes, that is,
\begin{equation}
\Tr(\rho^T E\rho E)=0\;.
\end{equation}
See Appendix~\ref{sec:SL2C}.  In this subsection we present our
understanding of these states.  We find that such a state has a
standard form described by one continuous complex parameter.

The spectral representation of $\rho$,
\begin{equation}
\rho=\sum_{i=1}^4\lambda_i\,\eta_i\eta_i^{\dag}\;,
\end{equation}
with four eigenvalues $\lambda_i>0$ and corresponding orthonormal
eigenvectors $\eta_i$, gives that
\begin{eqnarray}
\Tr(\rho^T E\rho E)
&\!\!\!=&\!\!\!
\sum_{i,j}\lambda_i\lambda_j
\Tr{\big[\eta_i^{\ast}\eta_i^T E
\eta_j\eta_j^{\dag} E\big]}
=\sum_{i,j}\lambda_i\lambda_j
{\big[\eta_i^T E\eta_j\big]
 \big[\eta_j^{\dag} E\eta_i^{\ast}\big]}
\nonumber\\
&\!\!\!=&\!\!\!
-\sum_{i,j}\lambda_i\lambda_j
{\big[\eta_i^T E\eta_j\big]
 \big[\eta_i^{\dag} E\eta_j^{\ast}\big]}
=-\sum_{i,j}\lambda_i\lambda_j\,
\big|\eta_i^T E\eta_j\big|^2\;.
\end{eqnarray}
Since the four eigenvalues are positive, the only way to have
$\Tr(\rho^T E\rho E)=0$ is to have all the eigenvectors orthogonal in
the scalar product defined by $E$,
\begin{equation}
\label{eq:etaiEetaj}
\eta_i^T E\eta_j=0
\qquad\mbox{for}\qquad
i,j=1,2,3,4\;.
\end{equation}
Since $E^T=-E$, this scalar product is antisymmetric,
$\psi^TE\phi=-\phi^TE\psi$, hence every vector is orthogonal to
itself, $\psi^TE\psi=0$.

Equation~(\ref{eq:etaiEetaj}) means that the antisymmetric scalar
product vanishes identically in the four dimensional subspace
$\Img\rho$ spanned by the eigenvectors $\eta_i$ with $i=1,2,3,4$.  We
may use this observation in order to search numerically for PPT states
of this type.

\subsubsection{A random search method}

It is a straightforward exercise to construct a random four
dimensional subspace $\mathcal{U}\subset\mathbbm{C}^8$ where the
antisymmetric scalar product vanishes identically.  We may start with
a random normalized vector $\psi_1$.  The two conditions
$\psi_1^{\dag}\psi=0$, $\psi_1^TE\psi=0$ restrict $\psi$ to a six
dimensional subspace, and we choose $\psi_2$ as a random normalized
vector in this subspace.  Next, the conditions $\psi_i^{\dag}\psi=0$,
$\psi_i^TE\psi=0$ for $i=1,2$ restrict $\psi$ to a four dimensional
subspace, and we choose $\psi_3$ as a random normalized vector in this
subspace.  Finally, we choose $\psi_4$ as a random normalized vector
in a two dimensional subspace.  In this way we get vectors
$\psi_1,\psi_2,\psi_3,\psi_4$ such that
\begin{equation}
\psi_i^{\dag}\psi_j=\delta_{ij}\;,\qquad
\psi_i^TE\psi_j=0\;.
\end{equation}
The vectors $\phi_i=E\psi_i^{\ast}$ for $i=1,2,3,4$ lie in the orthogonal
subspace $\mathcal{U}^{\perp}$, since
\begin{equation}
\phi_i^{\dag}\psi_j=-\psi_i^TE\psi_j=0\;,
\end{equation}
and they satisfy the same relations as the vectors $\psi_i$,
\begin{eqnarray}
\phi_i^{\dag}\phi_j&\!\!\!=&\!\!\!
-\psi_i^TE^2\psi_j^{\ast}=
(\psi_i^{\dag}\psi_j)^{\ast}=\delta_{ij}\;,
\nonumber\\
\phi_i^TE\phi_j&\!\!\!=&\!\!\!
-\psi_i^{\dag}E^3\psi_j^{\ast}=
(\psi_i^TE\psi_j)^{\ast}=0\;.
\end{eqnarray}

For a given subspace $\mathcal{U}$ we may try to construct numerically
a PPT state $\rho$ with $\Img\rho=\mathcal{U}$, by the method
described in Subsection~\ref{subsubsec:PPTstatesbisep}.  Again we find
that the construction fails in general, but we may select a suitable
subspace with vanishing antisymmetric scalar product where the
construction succeeds.

As remarked in Subsection~\ref{subsubsec:PPTstatesbisep}, when we
search for a random subspace $\mathcal{U}$ where we can find a matrix
$\rho$ satisfying all three biseparability
conditions~(\ref{eq:bisep1}), (\ref{eq:bisep2}),
and~(\ref{eq:bisep3}), disregarding the positivity conditions, then it
happens only in about one case out of eight that the $\rho$ we find is
positive.  Surprisingly, when we now do the same search restricted to
subspaces with the special property that the antisymmetric scalar
product vanishes identically, then the positivity conditions hold, not
every time, but almost every time.

\subsubsection{Explicit construction}

By studying our numerical examples of such states we find empirically
that they have certain properties which enable us to construct them
explicitly.  Again we have no strict proof that these properties are
necessary, but it is a very powerful working hypothesis to assume that
they hold.

Like before, we introduce three sets of basis vectors for the subspace
$\Img\rho$, $e_i=x_i\otimes u_i$, $f_j=y_j\otimes_sv_j$, and
$g_k=w_k\otimes z_k$, with $x_i,y_j,z_k\in\mathbbm{C}^2$ and
$u_i,v_j,w_k\in\mathbbm{C}^4$.

The vanishing of the antisymmetric scalar product in $\Img\rho$ means
that
\begin{equation}
e_i^TEe_j=(x_i^T\epsilon x_j)(u_i^T(\epsilon\otimes\epsilon)\,u_j)=0
\qquad\mbox{for}\qquad
i,j=1,2,3,4\;.  
\end{equation}
We have that $x_i^T\epsilon x_j=0$ for $i=j$, but
$x_i^T\epsilon x_j\neq 0$ for $i\neq j$.  Hence we must have
\begin{equation}
\label{eq:uieeuj}
u_i^T(\epsilon\otimes\epsilon)\,u_j=0
\qquad\mbox{for}\qquad
i\neq j\;.
\end{equation}
Note that $\epsilon$ is antisymmetric, but $\epsilon\otimes\epsilon$
is symmetric,
\begin{equation}
(\epsilon\otimes\epsilon)^T
=\epsilon^T\otimes\epsilon^T
=(-\epsilon)\otimes(-\epsilon)
=\epsilon\otimes\epsilon\;.
\end{equation}

In our numerical examples we see that equation~(\ref{eq:uieeuj}) is
solved in the following remarkable way.  The vectors $u_i$ with
$i=1,2,3,4$ can not all be product vectors, because then $\rho$ would
be separable.  Instead, each $u_i$ is a linear combination of two
product vectors, as follows,
\begin{equation}
\label{eq:ulincomb2}
u_i=a_{ikl}\,y_k\otimes z_l+a_{imn}\,y_m\otimes z_n\;.
\end{equation}
There is no sum here over the indices $k,l,m,n$, and we have always
$k\neq m$ and $l\neq n$.  For each value of $i$, six different
combinations occur for $klmn$, and we may order the vectors $y$ and
$z$ in such a way that we get the index combinations listed in
Table~\ref{tab:allowedind}.  Since
\begin{equation}
(y_k\otimes z_l)^T(\epsilon\otimes\epsilon)\,(y_m\otimes z_n)
=(y_k^T\epsilon\,y_m)\,(z_l^T\epsilon\,z_n)
=0
\quad\mbox{if}\quad
k=m
\quad\mbox{or}\quad
l=n
\;,
\end{equation}
we are guaranteed that $u_i$ and $u_j$ satisfy the orthogonality
relation in equation~(\ref{eq:uieeuj}) if
\begin{eqnarray}
u_i&\!\!\!=&\!\!\!a_{ikl}\,y_k\otimes z_l+a_{imn}\,y_m\otimes z_n\;,
\nonumber\\
u_j&\!\!\!=&\!\!\!a_{jkn}\,y_k\otimes z_n+a_{iml}\,y_m\otimes z_l\;.
\end{eqnarray}
We see from Table~\ref{tab:allowedind} that two vectors $u_i$ and
$u_j$ are always orthogonal in two ways.  For example,
\begin{eqnarray}
u_1&\!\!\!=&\!\!\!
 a_{112}\,y_1\otimes z_2+a_{121}\,y_2\otimes z_1
=a_{134}\,y_3\otimes z_4+a_{143}\,y_4\otimes z_3\;,
\nonumber\\
u_2&\!\!\!=&\!\!\!
 a_{211}\,y_1\otimes z_1+a_{222}\,y_2\otimes z_2
=a_{233}\,y_3\otimes z_3+a_{244}\,y_4\otimes z_4\;.
\end{eqnarray}

\begin{table}[h]
\centering
\begin{tabular}{|c|cccccc|}
\hline
$i$ & \multicolumn{6}{c|}{$klmn$}\\
\hline
1 & 1221 & 1331 & 1441 & 2332 & 2442 & 3443\\
2 & 1122 & 1342 & 1432 & 2341 & 2431 & 3344\\
3 & 1133 & 1243 & 1423 & 2134 & 2244 & 3241\\
4 & 1144 & 1234 & 1324 & 2143 & 2233 & 3142\\
\hline
\end{tabular}
\caption{The allowed index combinations in equation~(\ref{eq:ulincomb2}).}
\label{tab:allowedind}
\end{table}

We now transform $x,y,z$ to the standard form defined in
equation~(\ref{eq:standformxyz}), with complex parameters
$t_1,t_2,t_3$.  We find that the linear dependencies listed in
Table~\ref{tab:allowedind} require that $t_2=t_3$, and when this
relation holds they give a unique solution for $u$ depending on the
single complex parameter $t=t_2=t_3$.  The solution is
\begin{equation}
\label{eq:standformu}
u=
\begin{pmatrix}
 0 &  t &  t &  t\\
 1 &  0 &  1 & -t\\
-1 &  0 &  1 & -t\\
 0 &  1 & -1 & -1
\end{pmatrix}.
\end{equation}
The condition that the four vectors $g_k=w_k\otimes z_k$ must be
linear combinations of the vectors $e_i=x_i\otimes u_i$ requires that
also $t_1=t$.  We find the overall solution $v=w=u$, giving the
vectors
\begin{equation}
e=
\begin{pmatrix}
 0 &  0 &  t &  t^2\\
 1 &  0 &  1 & -t^2\\
-1 &  0 &  1 & -t^2\\
 0 &  0 & -1 & -t\\
 0 &  t & -t &  t\\
 0 &  0 & -1 & -t\\
 0 &  0 & -1 & -t\\
 0 &  1 &  1 & -1
\end{pmatrix},
\quad
f=
\begin{pmatrix}
 0 &  0 &  t &  t^2\\
 1 &  0 &  1 & -t^2\\
 0 &  t & -t &  t\\
 0 &  0 & -1 & -t\\
-1 &  0 &  1 & -t^2\\
 0 &  0 & -1 & -t\\
 0 &  0 & -1 & -t\\
 0 &  1 &  1 & -1
\end{pmatrix},
\quad
g=
\begin{pmatrix}
 0 &  0 &  t &  t^2\\
 0 &  t & -t &  t\\
 1 &  0 &  1 & -t^2\\
 0 &  0 & -1 & -t\\
-1 &  0 &  1 & -t^2\\
 0 &  0 & -1 & -t\\
 0 &  0 & -1 & -t\\
 0 &  1 &  1 & -1
\end{pmatrix}.
\end{equation}

Thus we arrive at the following explicit standard form for $\rho$,
depending on the single complex parameter $t$,
\begin{equation}
\label{eq:standformrhoinv0}
\rho
=a\sum_{i=1}^4\lambda_i\,e_ie_i^{\dag}
=a\sum_{i=1}^4\lambda_i\,f_if_i^{\dag}
=a\sum_{i=1}^4\lambda_i\,g_ig_i^{\dag}
\end{equation}
with
\begin{equation}
\lambda_1=|t|^2\,|1+t|^2\;,\qquad
\lambda_2=|1+t|^2\;,\qquad
\lambda_3=|t|^2\;,\qquad
\lambda_4=1\;,
\end{equation}
and
\begin{equation}
a=\frac{1}{5|t|^4+10|t|^2+1+(3|t|^2+1)\,|1+t|^2}\;.
\end{equation}

\subsubsection{The effect of partial transposition}

The effect of the partial transposition $T_1$, for example, on the PPT
state $\rho$ given in equation~(\ref{eq:standformrhoinv0}) is that
\begin{equation}
\rho^{T_1}
=a\sum_{i=1}^4\lambda_i\,\widetilde{e}_i\widetilde{e}_i^{\dag}\;,
\end{equation}
with $\widetilde{e}_i=x_i^{\ast}\otimes u_i$.  The two dimensional
vectors $x$ are complex conjugated while the four dimensional vectors
$u$ are unchanged.  If $x$ and $u$ have their standard forms as given
in equation~(\ref{eq:standformxyz}) (with $t_1=t$) and in
equation~(\ref{eq:standformu}), it means that the parameter $t$ is
complex conjugated in $x$ but not in $u$.

We find that the complex conjugation of the standard form of $u$ is
equivalent to a linear transformation on product form.  That is, there
exists a $2\times 2$ matrix $W$ such that
\begin{equation}
(W\otimes W)\,u_i=C_i\,u_i^{\ast}\;,
\end{equation}
with four complex normalization constants $C_i$ having different
phases but equal absolute values, $|C_i|=|C_j|$ for $i,j=1,2,3,4$.
There are four solutions for $W$, since the general solution
\begin{equation}
W
=\begin{pmatrix}
-\epsilon_1t^{\ast}(1-\epsilon_1\,|t|+\epsilon_2\,|1+t|)
&
|t|\,(t^{\ast}+\epsilon_1\,|t|)
\\
|t|+\epsilon_1t^{\ast}
&
|t|\,(1-\epsilon_1\,|t|+\epsilon_2\,|1+t|)
\end{pmatrix}
\end{equation}
contains two arbitrary signs
$\epsilon_1=\pm 1$ and $\epsilon_2=\pm 1$.

It follows when we define $V=\mathbbm{1}\otimes W\otimes W$, with the
$2\times 2$ unit matrix $\mathbbm{1}$, and introduce a normalization
factor $b$, that
\begin{equation}
b\,V\rho^{T_1}V^{\dag}=\rho^{\ast}=\rho^T\;.
\end{equation}
This shows that the partial transpose $\rho^{T_1}$ has a standard form
like the standard form of $\rho$, but with the complex conjugated
parameter value $t^{\ast}$.  The same observation applies of course
also to the partial transposes $\rho^{T_2}$ and $\rho^{T_3}$, and to
the total transpose $\rho^T=\rho^{\ast}$.

\section{Summary and outlook}

We have presented here a numerical survey of entangled PPT states in
the system of three qubits, with an emphasis on the extremal PPT
states.  Important characteristics of a state are the ranks of the
state itself and of its three single partial transposes.

For ranks equal to five or higher very few rank combinations are
missing, the variety could not be much larger than what we see.  This
is true also for the extremal PPT states, where there is an upper
bound of 193 for the square sum of ranks.  The existence of a large
variety of very different states indicates that it will not be easy to
understand all the extremal PPT states analytically, in the same way
as we understand the pure states.

A natural place to start a project to gain some analytical
understanding is with the rank 4444 states, which are the lowest rank
entangled PPT states.  Another good reason for our interest in these
states is that they have genuine tripartite entanglement, since they
are separable with respect to any bipartition.  An unanswered question
about the rank four states is why the rank combination 4444 is the
only one observed.

The most important advance reported here is that we have uncovered the
mathematical structure behind the rank 4444 entangled PPT states, and
shown how to reproduce analytically all the states found numerically.
They fall into two distinct classes with very different analytical
representations.  A sobering fact is that the known states of the UPB
type were not found numerically, although they appear as just a
special case in our classification scheme.  We take this as a reminder
that there may exist unknown types of rank four PPT states that we
have missed because they are not generic.

We intend to continue our project and try to understand for example
the rank 5555 extremal PPT states in the three qubit system.  The
numerical study of higher dimensional tripartite systems quickly
becomes impractical, simply because the dimensions grow rapidly.

\appendix

\section{Partial transpositions}
\label{sec:partialtranspositions}

Write an $8\times 8$ complex matrix as a $4\times 4$ matrix of
$2\times 2$ matrices,
\begin{equation}
X=
\begin{pmatrix}
A&B&C&D\\
E&F&G&H\\
I&J&K&L\\
M&N&O&P
\end{pmatrix}.
\end{equation}
The partial transposition $T_1$ moves $4\times 4$ submatrices,
\begin{equation}
X^{T_1}=
\begin{pmatrix}
A&B&I&J\\
E&F&M&N\\
C&D&K&L\\
G&H&O&P
\end{pmatrix}.
\end{equation}
$T_2$ moves $2\times 2$ submatrices within $4\times 4$ submatrices,
\begin{equation}
X^{T_2}=
\begin{pmatrix}
A&E&C&G\\
B&F&D&H\\
I&M&K&O\\
J&N&L&P
\end{pmatrix}.
\end{equation}
$T_3$ transposes the $2\times 2$ submatrices,
\begin{equation}
X^{T_3}=
\begin{pmatrix}
A^T&B^T&C^T&D^T\\
E^T&F^T&G^T&H^T\\
I^T&J^T&K^T&L^T\\
M^T&N^T&O^T&P^T
\end{pmatrix}.
\end{equation}
If $X$ is symmetric under all three partial transpositions, then it
has the form
\begin{equation}
X=
\begin{pmatrix}
A&B&C&D\\
B&F&D&H\\
C&D&K&L\\
D&H&L&P
\end{pmatrix}
\end{equation}
with $A^T=A$, $B^T=B$, and so on.  In particular, $X$ is symmetric,
$X^T=X^{T_1T_2T_3}=X$.  If in addition $X$ is Hermitian, then it is
real.

\section{$\mbox{SL}(2,\mathbbm{C})$, Lorentz transformations, and Lorentz invariants}
\label{sec:SL2C}

Let $\epsilon$ be the two dimensional Levi--Civita symbol,
\begin{equation}
\epsilon=\begin{pmatrix} 0&1\\ -1&0\end{pmatrix}.
\end{equation}
It is a square root of $-1$, $\epsilon^2=-\mathbbm{1}$.  The Lie group
$\mbox{SL}(2,\mathbbm{C})$ consists of all $2\times 2$ complex
matrices with unit determinant.  If $V\in\mbox{SL}(2,\mathbbm{C})$,
\begin{equation}
V=\begin{pmatrix} a&b\\ c&d\end{pmatrix},
\end{equation}
then
\begin{equation}
V^{-1}=\begin{pmatrix} d&-b\\ -c&a\end{pmatrix}
=-\epsilon V^T\epsilon\;.
\end{equation}
Thus $V\epsilon V^T=\epsilon$, in this sense $\epsilon$ is an
invariant tensor under $\mbox{SL}(2,\mathbbm{C})$ transformations.

A general $2\times 2$ Hermitian matrix may be written as
\begin{equation}
X=\begin{pmatrix}
x^0+x^3 & x^1-\rmi x^2\\
x^1+\rmi x^2 & x^0-x^3\end{pmatrix}
=x^{\mu}\sigma_{\mu}\;,
\end{equation}
where $x^{\mu}$ is a real fourvector, $\sigma_0=\mathbbm{1}$ is the
unit matrix, and $\sigma_j$ for $j=1,2,3$ are the Pauli matrices.  The
determinant of $X$ is
\begin{equation}
\det(X)
=-\frac{1}{2}\Tr(X^T\epsilon X\epsilon)
=g_{\mu\nu}x^{\mu}x^{\nu}\;,
\end{equation}
where $g_{\mu\nu}$ is the metric tensor,
\begin{equation}
g_{\mu\nu}=\begin{pmatrix}
1 & 0 & 0 & 0\\
0 &-1 & 0 & 0\\
0 & 0 &-1 & 0\\
0 & 0 & 0 &-1
\end{pmatrix}.
\end{equation}
More generally, for $X=x^{\mu}\sigma_{\mu}$ and
$Y=y^{\mu}\sigma_{\mu}$ we have
\begin{equation}
\Tr(X^T\epsilon Y\epsilon)
=-2g_{\mu\nu}x^{\mu}y^{\nu}
=-2x^{\mu}y_{\mu}\;.
\end{equation}
The transformation $X\mapsto\wt{X}=VXV^{\dag}$ with $\det V=1$ is a
continuous Lorentz transformation.  It leaves the determinant
invariant, and leaves the scalar product between two fourvectors
invariant because
$\det V^T=\det V^{\dag}=1$, hence
$V^T\epsilon V=V^{\dag}\epsilon V^{\ast}=\epsilon$ and
\begin{equation}
\Tr(\wt{X}^T\epsilon\wt{Y}\epsilon)
=\Tr(X^T(V^T\epsilon V)Y(V^{\dag}\epsilon V^{\ast}))
=\Tr(X^T\epsilon Y\epsilon)\;.
\end{equation}
The parity inversion $\wt{x}^2=-x^2$ takes the form $\wt{X}=X^T$ and
leaves the scalar product invariant, although it is not of the form
$\wt{X}=VXV^{\dag}$.

In
$\mathbbm{C}^8=\mathbbm{C}^2\otimes\mathbbm{C}^2\otimes\mathbbm{C}^2$
the antisymmetric tensor
\begin{equation}
E=\epsilon\otimes\epsilon\otimes\epsilon
=\begin{pmatrix}
 0 & 0 & 0 & 0 & 0 & 0 & 0 & 1\\
 0 & 0 & 0 & 0 & 0 & 0 &-1 & 0\\
 0 & 0 & 0 & 0 & 0 &-1 & 0 & 0\\
 0 & 0 & 0 & 0 & 1 & 0 & 0 & 0\\
 0 & 0 & 0 &-1 & 0 & 0 & 0 & 0\\
 0 & 0 & 1 & 0 & 0 & 0 & 0 & 0\\
 0 & 1 & 0 & 0 & 0 & 0 & 0 & 0\\
-1 & 0 & 0 & 0 & 0 & 0 & 0 & 0
\end{pmatrix}
\end{equation}
is invariant under $\mbox{SL}\otimes\mbox{SL}\otimes\mbox{SL}$
transformations, in the sense that $VEV^T=E$ when
$V=V_1\otimes V_2\otimes V_3$ and
$V_1,V_2,V_3\in\mbox{SL}(2,\mathbbm{C})$.

A general $8\times 8$ Hermitian matrix may be written as
\begin{equation}
  A=a^{\lambda\mu\nu}\sigma_{\lambda}\otimes\sigma_{\mu}\otimes\sigma_{\nu}
\end{equation}
with $4\times 4\times 4=64$ real coefficients
\begin{equation}
a^{\mu \nu \lambda}
=\frac{1}{8}\,
\Tr(A\,( \sigma_\mu\otimes\sigma_\nu\otimes\sigma_\lambda))\;.
\end{equation}
A product transformation of the form $\wt{A}=VAV^{\dag}$ with
$V=V_1\otimes V_2\otimes V_3$, as above, acts as three independent
continuous Lorentz transformations on the three Lorentz indices.  Note
that the partial transpositions are discrete Lorentz transformations,
since they are parity inversions.

For Hermitian matrices $A,B$ the quantity
\begin{equation}
\Tr(A^TEBE)
=-8g_{\lambda\alpha}g_{\mu\beta}g_{\nu\gamma}
a^{\lambda\mu\nu}b^{\alpha\beta\gamma}
=-8a^{\lambda\mu\nu}b_{\lambda\mu\nu}
\end{equation}
is real and invariant under the product transformations
$\wt{A}=VAV^{\dag}$, $\wt{B}=VBV^{\dag}$.  It is also invariant under
all three partial transpositions.  Note that each Lorentz index on a
tensor $a^{\lambda\mu\nu}$ represents its own subsystem and can
therefore only be contracted against the corresponding index on
the tensor $b^{\lambda\mu\nu}$.

A density matrix $\rho$ in dimension $2\times 2\times 2$ has one
quadratic Lorentz invariant
\begin{equation}
I_2
=\rho^{\mu \nu \lambda}\rho_{\mu \nu \lambda}
=-\frac{1}{8}\,\Tr(\rho^TE\rho E)\geq 0\;.
\end{equation}
For pure states $\rho_i=\psi_i\psi_i^{\dag}$ we have that
\begin{equation}
\Tr(\rho_i^TE\rho_j E)
={[\psi_i^T\!E\psi_j][\psi_j^{\dag}E\psi_i^{\ast}]}
=-{[\psi_i^T\!E\psi_j][\psi_i^{\dag}E\psi_j^{\ast}]}
=-|\psi_i^T\!E\psi_j|^2\;.
\end{equation}
The inequality $I_2\geq 0$ follows because $\rho$ is always a convex
combination of pure states,
\begin{equation}
\rho=\sum_i\lambda_i\,\psi_i\psi_i^{\dag}
\qquad\mbox{with}\qquad
\lambda_i>0\;,
\quad
\sum_i\lambda_i=1\;,
\end{equation}
and hence
\begin{equation}
\Tr(\rho^TE\rho E)
=-\sum_{i,j}\lambda_i\lambda_j\,|\psi_i^T\!E\psi_j|^2
\leq 0\;.
\end{equation}

There are five different fourth order invariants obtained by different
combinations of index contractions, but one of these is simply the
square of the second order invariant. The four new invariants can be
written as
\begin{eqnarray}
I_{41}
&\!\!\!=&\!\!\!
\rho^{\mu \nu \lambda}\,\rho_{\mu\nu\gamma}\,
\rho^{\alpha \beta\gamma}\,\rho_{\alpha \beta \lambda}\;,
\nonumber\\
I_{42}
&\!\!\!=&\!\!\!
\rho^{\mu \nu \lambda}\,\rho_{\mu\beta\lambda}\,
\rho^{\alpha \beta\gamma}\,\rho_{\alpha \nu \gamma}\;,
\nonumber\\
I_{43}
&\!\!\!=&\!\!\!
\rho^{\mu \nu \lambda}\,\rho_{\mu\beta\gamma}\,
\rho^{\alpha \beta\gamma}\,\rho_{\alpha \nu \lambda}\;,
\\
I_{44}
&\!\!\!=&\!\!\!
\rho^{\mu \nu \lambda}\,\rho_{\mu}^{\phantom{\mu}\beta\gamma}\,
\rho^{\alpha}_{\phantom{\alpha} \nu\gamma}\,\rho_{\alpha \beta \lambda}\;.
\nonumber
\end{eqnarray}

Note that all these Lorentz invariants are invariant under
$\mbox{SL}\otimes\mbox{SL}\otimes\mbox{SL}$ transformations of a
density matrix without subsequent normalization to unit trace.
Division of the fourth order invariants by the square of the second
order invariant gives true invariants that are also independent of the
normalization of the density matrix.  They may be used in order to
test whether two density matrices belong to the same
$\mbox{SL}\otimes\mbox{SL}\otimes\mbox{SL}$ equivalence class.

\section{Standard forms of sets of vectors}
\label{sec:standardforms}

Given four vectors $x_i\in\mathbbm{C}^2$, $i=1,2,3,4$.  We write
\begin{equation}
x=(x_1,x_2,x_3,x_4)
=\begin{pmatrix}a&c&e&g\\ b&d&f&h\end{pmatrix}.
\end{equation}
We consider here the generic case with $\det(x_i,x_j)\neq 0$ for
$i\neq j$.  Multiplication by the matrix
\begin{equation}
U=\begin{pmatrix}d&-c\\ -b&a\end{pmatrix}
\end{equation}
and a subsequent normalization of the vectors gives the form
\begin{equation}
y=\begin{pmatrix}1&0&1&t_2\\ 0&1&t_1&1\end{pmatrix},
\end{equation}
with $\det(y_i,y_j)\neq 0$ for $i\neq j$, which means that
$t_1t_2\neq 0,1$.  Multiplication by
\begin{equation}
V=\begin{pmatrix}-t_1&0\\ 0&1\end{pmatrix}
\end{equation}
and normalization now gives the standard form
\begin{equation}
\label{eq:standform1}
z=\begin{pmatrix}1&0&1&t\\ 0&1&-1&1\end{pmatrix},
\end{equation}
with one variable parameter $t=-t_1t_2\neq 0,-1$.  Since
\begin{equation}
t=-\frac{\det(z_1,z_3)\det(z_2,z_4)}{\det(z_1,z_4)\det(z_2,z_3)}\;,
\end{equation}
and this ratio of determinants is invariant under nonsingular linear
transformations and vector normalizations, we have that
\begin{equation}
t=-\frac{\det(x_1,x_3)\det(x_2,x_4)}{\det(x_1,x_4)\det(x_2,x_3)}
=-\frac{(af-be)(ch-dg)}{(ah-bg)(cf-de)}\;.
\end{equation}
We see that $t$ will be complex in the generic case.  In the special
case where $t$ is real and positive, we may multiply by
\begin{equation}
W=\begin{pmatrix}1&0\\ 0&\sqrt{t}\end{pmatrix}
\end{equation}
and normalize so as to obtain the standard form
\begin{equation}
w=\begin{pmatrix}1&0&1&\sqrt{t}\\ 0&1&-\sqrt{t}&1\end{pmatrix},
\end{equation}
where the vectors are real and pairwise orthogonal,
$w_i^{\dag}w_j=w_i^Tw_j=0$ for $i,j=1,2$ and $i,j=3,4$.

Instead of equation~(\ref{eq:standform1}) we might have chosen one of
two alternative standard forms,
\begin{equation}
\label{eq:standform2}
z'=\begin{pmatrix}1&1&0&t'\\ 0&-1&1&1\end{pmatrix},
\end{equation}
or
\begin{equation}
\label{eq:standform3}
z''=\begin{pmatrix}1&1&t''&0\\ 0&-1&1&1\end{pmatrix}.
\end{equation}
The invariant formula for $t$, equation~(\ref{eq:standform1}), gives
that
\begin{equation}
t=-1-t'\;,\qquad t=-\frac{1}{1+t''}\;,
\end{equation}
or inversely,
\begin{equation}
t'=-1-t\;,\qquad t''=-1-\frac{1}{t}\;.
\end{equation}
In the case where $t$ is real we see that $t>0$ gives $t'<0$, $t''<0$,
whereas $-1<t<0$ gives $t'<0$, $t''>0$, and $t<-1$ gives $t'>0$,
$t''<0$.  Thus, if $t$ is real and $t\neq 0,-1$, there is always
exactly one pairing of the four vectors, either $(x_1x_2)(x_3x_4)$,
$(x_1x_3)(x_2x_4)$, or $(x_1x_4)(x_2x_3)$, such that there exists a
linear transformation which will make both pairs real and orthogonal.

\section{Unextendible product bases}
\label{sec:UPBs}

An unextendible product basis (a UPB) in a subspace $\mathcal{U}$ of
$\mathbbm{C}^2\otimes\mathbbm{C}^2\otimes\mathbbm{C}^2$ is a set of
orthogonal product vectors $e_i=x_i\otimes y_i\otimes z_i$ spanning
$\mathcal{U}$ such that no product vector is orthogonal to them all.
We assume that the vectors $x_i,x_j\in\mathbbm{C}^2$ are linearly
independent when $i\neq j$, and similarly with $y_i,y_j$ and
$z_i,z_j$.

Obviously, in order to have $e_i\perp e_j$ we must have either
$x_i\perp x_j$, $y_i\perp y_j$, or $z_i\perp z_j$.  These conditions
have a solution with four product vectors, unique up to permutations,
as illustrated in Figure~\ref{fig1}.  In a similar way as discussed in
Appendix~\ref{sec:standardforms} the vectors may always be transformed
to the real standard forms
\begin{equation}
\label{eq:standform4}
x=\begin{pmatrix}1&0&c_1&s_1\\ 0&1&-s_1&c_1\end{pmatrix},\quad
y=\begin{pmatrix}1&c_2&0&s_2\\ 0&-s_2&1&c_2\end{pmatrix},\quad
z=\begin{pmatrix}1&c_3&s_3&0\\ 0&-s_3&c_3&1\end{pmatrix},
\end{equation}
where $c_i=\cos\theta_i$, $s_i=\sin\theta_i$, and
$\theta_1,\theta_2,\theta_3$ are three angular parameters.

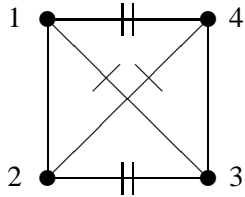
\begin{figure}[h]
\begin{center}
\begin{picture}(80,80)(0,0)
\put(10,10){\circle*{6}}
\put(10,70){\circle*{6}}
\put(70,10){\circle*{6}}
\put(70,70){\circle*{6}}
\put(10,10){\line( 1, 0){60}}
\put(10,70){\line( 1, 0){60}}
\put(10,10){\line( 0, 1){60}}
\put(70,10){\line( 0, 1){60}}
\put(10,10){\line( 1, 1){60}}
\put(10,70){\line( 1,-1){60}}
\put(27,43){\line( 1, 1){10}}
\put(53,43){\line(-1, 1){10}}
\put(38,64){\line( 0, 1){12}}
\put(42,64){\line( 0, 1){12}}
\put(38, 4){\line( 0, 1){12}}
\put(42, 4){\line( 0, 1){12}}
\put(-5,67){1}
\put(-5, 7){2}
\put(78, 7){3}
\put(78,67){4}
\end{picture}
\end{center}
\caption{Orthogonality relations of an unextendible product basis
$e_i=x_i\otimes y_i\otimes z_i$, $i=1,2,3,4$.
An unmarked line: $x_i\perp x_j$.
A line with one tick mark: $y_i\perp y_j$.
A line with two tick marks: $z_i\perp z_j$.}
\label{fig1}
\end{figure}

\section{A problem of finding product vectors in a subspace}
\label{sec:prodvecinsubspace}

Given four vectors $\psi_j\in\mathbbm{C}^8$ with components
$\psi_{ij}$, $i=1,\ldots,8$, $j=1,\ldots,4$.  We write a linear combination
of them as a matrix product,
\begin{equation}
\phi=\sum_{j=1}^4\alpha_j\psi_j=\psi\alpha
\end{equation}
with $\alpha\in\mathbbm{C}^4$.  Assume that $\phi$ is a tensor product
\begin{equation}
\label{eq:pv3}
\phi=
\begin{pmatrix} a\\ b\end{pmatrix}\otimes
\begin{pmatrix} c\\ d\end{pmatrix}\otimes
\begin{pmatrix} e\\ f\end{pmatrix}
=
\begin{pmatrix}
ace\\
acf\\
ade\\
adf\\
bce\\
bcf\\
bde\\
bdf
\end{pmatrix}
.
\end{equation}
The presence of the first factor in the tensor product implies the
equation
\begin{equation}
\label{eq:geneveqI}
(A-\mu B)\alpha=0\;,
\end{equation}
where $\mu=a/b$, and $A$ and $B$ are the following $4\times 4$
matrices,
\begin{equation}
A=
\begin{pmatrix}
\psi_{11} & \psi_{12} & \psi_{13} & \psi_{14}\\
\psi_{21} & \psi_{22} & \psi_{23} & \psi_{24}\\
\psi_{31} & \psi_{32} & \psi_{33} & \psi_{34}\\
\psi_{41} & \psi_{42} & \psi_{43} & \psi_{44}
\end{pmatrix},
\qquad
B=
\begin{pmatrix}
\psi_{51} & \psi_{52} & \psi_{53} & \psi_{54}\\
\psi_{61} & \psi_{62} & \psi_{63} & \psi_{64}\\
\psi_{71} & \psi_{72} & \psi_{73} & \psi_{74}\\
\psi_{81} & \psi_{82} & \psi_{83} & \psi_{84}
\end{pmatrix}
.
\end{equation}
The presence of the second factor implies the equation
\begin{equation}
\label{eq:geneveqII}
(C-\mu D)\alpha=0\;,
\end{equation}
where $\mu=c/d$, and where
\begin{equation}
C=
\begin{pmatrix}
\psi_{11} & \psi_{12} & \psi_{13} & \psi_{14}\\
\psi_{21} & \psi_{22} & \psi_{23} & \psi_{24}\\
\psi_{51} & \psi_{52} & \psi_{53} & \psi_{54}\\
\psi_{61} & \psi_{62} & \psi_{63} & \psi_{64}
\end{pmatrix},
\qquad
D=
\begin{pmatrix}
\psi_{31} & \psi_{32} & \psi_{33} & \psi_{34}\\
\psi_{41} & \psi_{42} & \psi_{43} & \psi_{44}\\
\psi_{71} & \psi_{72} & \psi_{73} & \psi_{74}\\
\psi_{81} & \psi_{82} & \psi_{83} & \psi_{84}
\end{pmatrix}
.
\end{equation}
The presence of the third factor implies the equation
\begin{equation}
\label{eq:geneveqIII}
(E-\mu F)\alpha=0\;,
\end{equation}
where $\mu=e/f$, and where
\begin{equation}
E=
\begin{pmatrix}
\psi_{11} & \psi_{12} & \psi_{13} & \psi_{14}\\
\psi_{31} & \psi_{32} & \psi_{33} & \psi_{34}\\
\psi_{51} & \psi_{52} & \psi_{53} & \psi_{54}\\
\psi_{71} & \psi_{72} & \psi_{73} & \psi_{74}
\end{pmatrix},
\qquad
F=
\begin{pmatrix}
\psi_{21} & \psi_{22} & \psi_{23} & \psi_{24}\\
\psi_{41} & \psi_{42} & \psi_{43} & \psi_{44}\\
\psi_{61} & \psi_{62} & \psi_{63} & \psi_{64}\\
\psi_{81} & \psi_{82} & \psi_{83} & \psi_{84}
\end{pmatrix}
.
\end{equation}

Equation~(\ref{eq:geneveqI}) is a generalized eigenvalue equation,
having in the generic case four different complex eigenvalues $\mu_i$
with corresponding eigenvectors $\alpha_i$, defining four vectors that
are product vectors in dimension $2\times 4$ of the form
\begin{equation}
\phi_i=\psi\alpha_i=x_i\otimes u_i
\qquad\mbox{with}\qquad
x_i=\begin{pmatrix} \mu_i \\ 1\end{pmatrix},
\quad
u_i=B\alpha_i\;.
\end{equation}

Similarly, equation~(\ref{eq:geneveqII}) gives four eigenvalues
$\mu_i$ with corresponding eigenvectors $\alpha_i$, defining four
vectors that are product vectors when we use the split tensor product
defined in Appendix~\ref{sec:splittensorproduct},
\begin{equation}
\phi_i=\psi\alpha_i=y_i\otimes_sv_i
\qquad\mbox{with}\qquad
y_i=\begin{pmatrix} \mu_i \\ 1\end{pmatrix},
\quad
v_i=D\alpha_i\;.
\end{equation}

Finally, equation~(\ref{eq:geneveqIII}) gives four eigenvalues
$\mu_i$ with corresponding eigenvectors $\alpha_i$, defining four
vectors that are product vectors in dimension $4\times 2$,
\begin{equation}
\phi_i=\psi\alpha_i=w_i\otimes z_i
\qquad\mbox{with}\qquad
z_i=\begin{pmatrix} \mu_i \\ 1\end{pmatrix},
\quad
w_i=F\alpha_i\;.
\end{equation}

If the two equations~(\ref{eq:geneveqI}) and~(\ref{eq:geneveqIII})
have a common eigenvector $\alpha$, then $\phi=\psi\alpha$ is a product
vector in two ways, both in dimension $2\times 4$ and in $4\times 2$.
This means that it is a product vector in dimension
$2\times 2\times 2$, and the same $\alpha$ is an eigenvector of
equation~(\ref{eq:geneveqII}).

Note that the standard method for solving the generalized eigenvalue
equations~(\ref{eq:geneveqI}), (\ref{eq:geneveqII}),
and~(\ref{eq:geneveqIII}) depends on the nonsingularity of the matrices
$B$, $D$, and $F$.  If one or more of these matrices are singular, we
may usually avoid the problem simply by making a random product
transformation $\psi\mapsto\widetilde{\psi}=V\psi$ with
$V=V_1\otimes V_2\otimes V_3$, then solving the problem with
$\widetilde{\psi}$ instead of $\psi$ and transforming back in the end.

\section{The split tensor product}
\label{sec:splittensorproduct}

We find it useful to define a split tensor product
so as to be able to take out the middle factor in a tensor product
of three factors.  Thus we define
\begin{equation}
x\otimes y\otimes z=y\otimes_s(x\otimes z)\;.
\end{equation}
For $y=(c,d)^T\in\mathbbm{C}^2$ and
$v=(p,q,r,s)^T\in\mathbbm{C}^4=\mathbbm{C}^2\otimes\mathbbm{C}^2$ we
define
\begin{equation}
y\otimes_sv
=
\begin{pmatrix}
cp\\
cq\\
dp\\
dq\\
cr\\
cs\\
dr\\
ds
\end{pmatrix}.
\end{equation}
This corresponds to equation~(\ref{eq:pv3}) with $p=ae$, $q=af$,
$r=be$, $s=bf$.



\pagebreak



\end{document}